Mid-infrared optical parametric amplifier using silicon nanophotonic waveguides

Xiaoping Liu[1], Richard M. Osgood, Jr.[1], Yurii A. Vlasov[2] & William M. J. Green[2]

All-optical signal processing is envisioned as an approach to dramatically decrease power consumption and speed up performance of next-generation optical telecommunications networks[1-3]. Nonlinear optical effects, such as four-wave mixing (FWM) and parametric gain, have long been explored to realize all-optical functions in glass fibers[4]. An alternative approach is to employ nanoscale engineering of silicon waveguides to enhance the optical nonlinearities by up to five orders of magnitude[5], enabling integrated chip-scale all-optical signal processing. FWM within silicon nanophotonic waveguides has recently been used to demonstrate several telecom-band ($\lambda\sim1550$nm) all-optical functions, including wavelength conversion[6-9], signal regeneration[10], and tunable optical delay[11]. Despite these important advances, strong two-photon absorption[12] (TPA) of the telecom-band pump has been a fundamental and unavoidable obstacle, limiting parametric gain to values on the order of a few dB[13]. Here we demonstrate a silicon nanophotonic optical parametric amplifier exhibiting gain as large as 25.4 dB, by operating the pump in the mid-IR near one-half the band-gap energy ($E\sim0.55$eV, $\lambda\sim2200$nm), at which parasitic TPA-related absorption vanishes[12,14]. This gain is high enough to compensate all insertion losses, resulting in 13 dB net off-chip amplification. Furthermore, dispersion engineering[15] dramatically increases the gain bandwidth to more than 220 nm, all realized using an ultra-compact 4 mm silicon chip. Beyond its significant relevance to all-optical signal processing, the broadband parametric gain also facilitates the simultaneous generation of multiple on-chip mid-IR sources through cascaded FWM, covering a 500 nm spectral range. Together, these results provide a foundation for the construction of silicon-based room-temperature mid-IR light sources including tunable chip-scale parametric oscillators[16-18], optical frequency combs[19], and supercontinuum generators[20]. In this manner, silicon nanophotonic technology may be extended to an entirely new class of mid-IR applications[21,22] including biochemical detection, environmental monitoring, and free-space communication, in which an integrated on-chip platform can be of great benefit.

[1]Department of Electrical Engineering, Columbia University, 1300 S. W. Mudd Building, 500 W. 120th Street, New York, NY 10027, USA.
[2]IBM Thomas J. Watson Research Center, Yorktown Heights, NY 10598, USA.

In order to achieve large, broadband nonlinear parametric gain in silicon, the FWM process[23] relies upon precise phase matching between the co-propagating pump, signal, and idler waves, as well as the presence of a strong nonlinear interaction, described by the effective nonlinearity parameter $\gamma$. In addition, the intrinsic nonlinear figure of merit of silicon FOM = $n_2/(\beta_{TPA} \cdot \lambda)$, where $n_2$ is the intensity-dependent refractive index and $\beta_{TPA}$ is the TPA coefficient, should be as large as possible. While a large TPA coefficient restricts FOM to values less than 0.4 near $\lambda$ = 1550 nm, at mid-IR wavelengths approaching silicon's TPA threshold at $\lambda \sim 2200$ nm, FOM can increase as much as ten-fold to FOM > 4. Therefore, the mid-IR is a highly promising spectrum for realizing chip-scale silicon nanophotonic all-optical devices whose performance is no longer restricted by parasitic nonlinear absorption.

The essential dispersion and nonlinearity characteristics for a silicon nanophotonic waveguide operating in the mid-IR were engineered through a comprehensive multi-parameter design space study, encompassing the nanoscale dimensions of the silicon core, the refractive index of the cladding, and the waveguide mode polarization (see Methods and Supplementary Information).

The resulting geometry ultimately selected for fabrication is shown in the SEM cross-section image in Fig. 1a. The silicon waveguide core has dimensions of 700 nm x 425 nm, and is designed to operate in the fundamental quasi-TM mode near λ = 2200 nm, as illustrated by the overlaid mode field profile. The nanoscale-engineered optical structure is precisely controlled using a combination of silicon epitaxy, deep-UV lithography, reactive ion etching, and dielectric deposition, within an advanced CMOS fabrication facility (see Methods).

The calculated dispersion coefficient $D$ and effective nonlinearity parameter $\gamma$ for this specific silicon nanophotonic waveguide configuration are plotted in Fig. 1b, illustrating the mid-IR spectral dependence near silicon's TPA threshold. The dispersion is designed to be zero at a wavelength of λ = 2260 nm. The pump wavelength is chosen at λ = 2170 nm such that the nanophotonic waveguide possesses anomalous dispersion conditions with $D$ = 1000 ps/nm/km, as required for broadband phase matching[23]. Strong optical confinement of the quasi-TM mode to the ~0.3um$^2$ silicon core results in a large effective nonlinearity of $\gamma$ = 110 (W·m)$^{-1}$ at λ = 2170 nm, a value more than ten times larger than that of planar chalcogenide glass waveguides[24], and four orders of magnitude larger than that of highly nonlinear optical fiber[4,25].

The mid-IR nonlinear characteristics of a 4 mm-long silicon nanophotonic waveguide are studied by simultaneously injecting picosecond pump pulses centered at λ = 2170 nm along with a cw tunable mid-IR laser signal, and observing the resulting FWM at the waveguide output (experimental details in Methods). The peak coupled pump power at the waveguide input is $P_p$ ~ 27.9 W, while the input signal power $P_{sig}$ is kept below 0.45 mW. For each spectrum shown in Fig. 1c, two photons from the high-intensity pulsed pump mix with a single photon from the tunable cw signal to generate a wavelength-converted idler photon, at a wavelength dictated by energy conservation[23]. The FWM process also produces an additional photon at the signal wavelength, which can ultimately contribute to signal amplification. Due to the pulsed nature of the pump, the generated idler and amplified signal also occur as short pulses.

The FWM spectra in Fig. 1c are analyzed to extract the on-chip parametric signal gain and idler conversion gain values plotted in Fig. 1d (see Methods). The experimental data reveals that the engineered silicon nanophotonic waveguide successfully functions as a mid-IR optical parametric amplifier (OPA), with very large maximum signal and idler gain values of 24 dB and 25.4 dB, respectively. Moreover, the on-chip parametric gain is significant enough to overcome substantial fiber-chip coupling losses (blue curve), demonstrating net off-chip gain as large as 10 dB for the signal and 13 dB for the idler. The overall on-chip gain bandwidth spans the range from 2060 nm to 2280 nm, with the two prominent gain peaks coinciding with the envelopes of the pulsed signal and idler spectra in Fig. 1c. The maximum parametric gain obtained using our ultra-compact 4 mm device is more than 100 times larger than previously observed in a 17 mm-long silicon waveguide pumped near $\lambda$ = 1550 nm, where gain has been limited to ~3 dB owing to strong TPA-induced saturation effects[13]. By operating the pump at longer wavelengths in the mid-IR, we have realized a chip-scale silicon nanophotonic OPA which breaks the limits set by TPA nonlinear absorption.

Additional experiments were carried out with the input signal tuned to the gain peak at $\lambda$ = 2238 nm, in order to thoroughly characterize the OPA's input-output and gain saturation characteristics. Figure 2a plots the output peak signal and idler power vs. input signal power, with pump power $P_p \sim 27.9$ W. The dashed fit lines highlight linear amplification over a dynamic range significantly larger than 30 dB, with signal/idler gain of 23.4 dB/24.7 dB. Figure 2b plots on-chip signal/idler gain vs. input pump power, with signal power $P_{sig} \sim 0.39$ mW. The intersection with the dotted line illustrates that on-chip transparency is reached with a pump power near $P_p \sim 7.5$ W. These pump requirements may be reduced by increasing the waveguide length, or by suitable nanoscale engineering of the waveguide cross-section to increase the effective nonlinearity. Figure 2b also shows that saturation does not begin until $P_p \sim 19$ W, at which point signal/idler gain both exceed 20 dB. The measurements reveal that this gain saturation is strongly correlated with self-limiting of the transmitted pump power, as shown by the black squares in Fig. 2c. Numerical models used to assess the origin of this self-limiting (see Methods) reveal that a residual TPA coefficient[12] of $\beta_{TPA} = 0.106$ cm/GW cannot alone account for the observed saturation at high input pump power (blue dash-dot curve). The significantly reduced value of $\beta_{TPA}$ at $\lambda = 2170$ nm, approximately seven times smaller than at $\lambda = 1550$ nm, suggests that higher-order nonlinear absorption dominates under these OPA operating conditions. The red dashed curve in Fig. 2c shows that when a three-photon absorption (3PA) coefficient[26] of $\gamma_{3PA} = 0.025$ cm$^3$/GW$^2$ is included in the simulation along with TPA, the model accurately predicts the experimental trend. However, even though 3PA is expected to play a role up to $\lambda \sim 3300$ nm, the characteristics of the OPA demonstrated here clearly illustrate that nonlinear absorption no longer presents a significant obstacle to obtaining parametric gain large enough for practical applications in all-optical signal processing[1-3].

The large on-chip parametric gain produces amplified picosecond signal and idler pulses having peak power of several hundred milliwatts within the silicon nanophotonic OPA, which can themselves mix with the pump to efficiently generate new wavelengths through cascaded FWM. As shown by the spectrum in Fig. 3, four orders of cascaded mixing products (numbered peaks 3-9) spanning a spectral bandwidth of 500 nm are observed when the pump is operated at λ = 2180 nm ($P_p$ ~ 21.8 W), and the signal is tuned to λ = 2240 nm with an increased input power of $P_{sig}$ ~ 4.4 mW. While on-chip parametric gain under these conditions is restricted to the spectral range 2086 nm < λ < 2273 nm (hatched region), an analysis of the wavelength conversion efficiency at each cascaded mixing peak can assist in further quantifying the OPA characteristics within the regions which experience no gain. For example, the various FWM combinations which can give rise to peak 5 are listed in Table 1. As depicted by the inset energy diagram in Fig. 3, the most sensible dominant term originates from non-degenerate FWM of peaks 1 and 3 as pumps, leading to wavelength conversion of a signal at peak 4 into an idler at peak 5. The conversion efficiency for this mixing term is approximately -12 dB. Applying a similar approach, the conversion efficiencies for peaks 6-9 range from -4 dB to -9 dB, values approximately 8-10 dB larger than those observed in recent silicon mid-IR wavelength conversion experiments[27,28]. These results illustrate that even far outside the parametric gain bandwidth, the present silicon nanophotonic OPA design can be used as a power-efficient generator of broadband mid-IR white light and/or multi-line sources.

Taken together, the results above provide strong encouragement for a broad extension of the capabilities of the integrated silicon nanophotonic platform into the mid-IR spectrum. Making use of the ultra-compact high-gain OPA demonstrated here, a direction of immediate interest is the introduction of resonant feedback to develop a silicon-based, low-threshold, highly portable optical parametric oscillator[16] (OPO). Novel broadband room-temperature light sources such as these would have wide-ranging applications in mid-IR spectroscopy, sensing, and free-space communication[21,22].

**Methods:**

**Dispersion and effective nonlinearity design calculations**

Extensive numerical modeling was performed in order to identify the physical structure of the optimal silicon nanophotonic waveguide possessing the dispersion and effective nonlinearity characteristics required for strong, phase-matched FWM effects within an ultra-compact millimeter-scale device. The waveguide design focused upon achieving the desired dispersion and effective nonlinearity characteristics for the fundamental quasi-TM-polarized mode rather than the quasi-TE mode, on account of the reduced sensitivity to scattering loss from sidewall roughness. Contour plots of the fundamental quasi-TM mode's zero-dispersion wavelength (ZDWL) and the effective nonlinearity parameter $\gamma$ at the ZDWL, as a function of the silicon core cross-sectional dimensions (645 nm < width < 735 nm; 380 nm < height < 470 nm) are included in the Supplementary Information, along with additional description of the design procedure.

**Silicon nanophotonic waveguide fabrication**

The silicon nanophotonic waveguides were fabricated at the IBM Microelectronics Research Laboratory, using 200 mm silicon-on-insulator (SOI) wafers from SOITEC. The wafers had a 3 μm buried oxide (BOX) layer, and a 250 nm silicon layer with a resistivity of 10 Ω·cm. In order to reach the 425 nm target SOI thickness established by the above design procedure, an additional 175 nm layer of undoped single-crystal silicon was epitaxially grown on top of the substrate. A 100 nm thick $SiO_2$ hardmask was then deposited by LPCVD. Nanophotonic waveguide patterns of variable width were exposed into a layer of resist by dose-striping, using 248 nm deep-UV lithography. The patterns were then transferred into the $SiO_2$ hardmask by RIE. After stripping the resist, the waveguides were etched into the SOI layer using a HBr-based RIE chemistry. A layer of $SiO_xN_y$ with n ~ 1.56 was then deposited by PECVD and partially planarized. Finally, a 3 μm thick $SiO_2$ film was deposited as the upper cladding. Optical access was accomplished by means of facets cleaved through the silicon waveguides.

**Waveguide characterization and FWM measurements**

Silicon nanophotonic waveguide propagation losses near λ = 2200 nm were measured using the cutback method, by measuring transmission of broadband amplified spontaneous emission (ASE) from the tunable mid-IR cw laser (Photonics Innovations SFTL) through waveguides having various lengths from 4 - 45 mm. Unpolarized emission from the laser operating below threshold was coupled into a single-mode fibre using a microscope objective and micropositioner stage. Lensed tapered fibers were aligned to the silicon nanophotonic waveguide facets with x-y-z piezoelectric positioners for on-/off-chip coupling. Transmission spectra at the waveguide output were characterized with a mid-IR optical spectrum analyzer (Yokogawa AQ6375) operating at 1 nm resolution bandwidth. The propagation loss varied from approximately 4 dB/cm at 2030 nm, to 10 dB/cm at 2500 nm. Subtracting the linear propagation loss from the total measured fiber-to-fiber insertion loss, a coupling loss from lensed fibre into the silicon nanophotonic waveguide of approximately 6.5 +/- 1 dB/facet was estimated.

For the FWM experiments, collimated beams from the mid-IR picosecond pulsed pump (Coherent Mira-OPO, FWHM ~ 2 ps, repetition rate = 76 MHz) and cw tunable signal lasers first passed through variable neutral density filters for power control, and were then coupled into separate single-mode fibres. The pump and signal were multiplexed using a fused fiber directional coupler, (99/1 ratio for parametric gain measurements, 90/10 ratio for cascaded FWM measurements). The combined optical power was coupled into a 4 mm long waveguide using lensed tapered fibres. Polarization controllers were used to co-polarize both pump and signal to selectively excite the fundamental quasi-TM waveguide mode.

**Extraction of parametric conversion efficiency and signal gain**

The peak power of the converted idler pulse at the output of the silicon nanophotonic waveguide, $P_{idler\_out}$, was derived from the measured FWM spectra, according to $P_{idler\_out} = F\left(\int P_{idler\_avg}(\lambda)d\lambda\right)$. In order to convert the time-averaged idler power $P_{idler\_avg}$ measured by the OSA into peak power, the spectrally integrated power was weighted by the duty cycle factor $F = 1/(76 \text{ MHz}*2 \text{ ps})$, due to the pulsed nature of the experiment. A similar procedure was applied to find the pulsed signal output power $P_{signal\_out}$. A 2 nm wide band-stop filter was first numerically applied to the time-averaged signal spectrum, in order to exclude the power remaining in the narrowband cw tone. The peak signal power was then computed according to $P_{signal\_out} = F\left(\int P_{signal\_filtered\_avg}(\lambda)d\lambda\right)$. Finally, to find the cw signal power at the waveguide input $P_{signal\_in}$, the output cw signal power was measured (with the pump off) and corrected to account for total propagation losses of $\alpha$ dB incurred through the 4 mm long device, $P_{signal\_in} = 10^{\alpha/10}\left(\int P_{signal\_out\_pump\_off}(\lambda)d\lambda\right)$. Using the above quantities, the on-chip idler conversion gain $\eta$ was then defined as the ratio of peak idler power and input cw signal power, $\eta = P_{idler\_out}/P_{signal\_in}$. Accordingly, on-chip signal gain was given by $G = P_{signal\_out}/P_{signal\_in}$.

The error bars in the on-chip parametric gain data were calculated according to uncertainty in the measured propagation and facet coupling losses, as well as to account for the contribution of the OSA noise floor accumulated when integrating the signal/idler power at the waveguide output.

**Numerical simulations of pulse propagation and TPA/3PA coefficient fitting**

A perturbed nonlinear Schrödinger equation (NLSE) model[5] was used to simulate the pulse dynamics in the silicon nanophotonic waveguide. The underlying NLSE was solved with a split-step Fourier transform technique, where the linear part of the NLSE was solved in the Fourier domain, and the nonlinear part was solved in the time domain with a $4^{th}$-order Runge-Kutta method. Nonlinear loss mechanisms including TPA, 3PA, and related free-carrier absorption (FCA) were taken into account. The TPA and 3PA coefficients were varied within the range of experimental uncertainty in the bulk silicon data[12,26] in order to best fit the measured pump self-limiting curve. All other parameters, including waveguide dispersion, effective nonlinearity, propagation loss, and input power, were matched to the corresponding experimental conditions.

**End Notes:**

**Acknowledgements**

The authors gratefully acknowledge the staff at the IBM Microelectronics Research Laboratory where the silicon nanophotonic OPA devices were fabricated, K. Reuter and B. Price for assistance with SEM images, and J. Dadap for assistance with optimization of the pulsed laser system used in the experiments. They also recognize T. Kippenberg, G. Roelkens, and R. Soref for many helpful and motivating discussions.

**Author information**

Correspondence and requests for materials should be addressed to W.M.J.G. (wgreen@us.ibm.com).

**Tables:**

**Table 1 | Mixing terms contributing to peak 5 of cascaded FWM spectrum shown in Fig. 3.**

| Mixing combination (idler: pump 1, pump 2, signal) | Effective pump power (dBm) | Signal power (dBm) | Idler power (dBm) | Conversion efficiency (dB) |
|---|---|---|---|---|
| 5:3,1,4 | 32.1 | 23.7 | 11.2 | -12.5 |
| 5:3,3,1 | 25.6 | 38.5 | 11.2 | -27.3 |
| 5:3,1,2 | 32.1 | 3.7 | 11.2 | 7.5 |
| 5:1,1,6 | 38.5 | 7.2 | 11.2 | 4.0 |
| 5:1,4,8 | 31.1 | -6.2 | 11.2 | 17.4 |

All possible degenerate and non-degenerate FWM combinations giving rise to peak 5 in Fig. 3 are listed in the first column. These combinations are labeled according to the role of each numbered peak (idler: pump 1, pump 2, signal) in the given FWM process. The power associated with each peak in the cascaded FWM spectrum is evaluated as described in Methods. The effective pump power is given by the geometric average of the power in each pump, $\sqrt{P_{pump1}P_{pump2}}$. The conversion efficiency is defined as the ratio of the idler power and the signal power, $P_{idler}/P_{signal}$. Mixing terms which appear to suggest a conversion efficiency greater than zero, i.e. idler conversion gain, are ruled out on account of the fact that peak 5 lies well outside the experimentally determined parametric gain bandwidth. Of the remaining mixing combinations, the dominant source for peak 5 is likely to originate from the combination with the largest effective pump power, i.e. combination 5:3,1,4.

**Figure legends:**

**Figure 1 | Engineered silicon nanophotonic waveguide characteristics, mid-IR FWM experiments, and broadband on-chip optical parametric amplification. a**, Scanning electron microscope cross-section image of the silicon nanophotonic waveguide, consisting of a silicon core 700 nm wide with a height of 425 nm, surrounded by an upper and lower cladding of $SiO_2$ (n ~ 1.46), and a lateral cladding layer of $SiO_xN_y$ (n ~ 1.56). The $E_y$ component of the fundamental quasi-TM mode at λ = 2200 nm is shown as the overlaid colour map. **b**, Simulated dispersion coefficient *D* (blue curve) and effective nonlinearity parameter γ (red curve). The waveguide dimensions are engineered for zero dispersion at λ = 2260 nm, while providing anomalous dispersion characteristics at the pump wavelength (*D* = 1000 ps/nm/km at λ = 2170 nm) as required for phase matching. The effective nonlinearity parameter γ has a value of γ = 110 (W·m)$^{-1}$ at the pump wavelength. **c**, Series of FWM spectra taken after propagation of the pump and signal through the 4 mm-long silicon nanophotonic waveguide. The pulsed pump (FWHM ~ 2 ps, repetition rate = 76 MHz, $P_p$ ~ 27.9 W) is centered at λ = 2170 nm, while the cw signal ($P_{sig}$ < 0.45 mW) is scanned stepwise through the range 2193 – 2325 nm, generating corresponding idler terms from 2141 – 2028 nm. **d**, On-chip parametric signal gain (black triangles) and idler conversion gain (red circles), exhibiting broadband amplification across ~ 220 nm, with peak values greater than 25 dB. Net off-chip gain is obtained where on-chip gain exceeds fiber-chip insertion losses (blue line), with a maximum value of 13 dB.

**Figure 2 | Mid-IR optical parametric amplifier performance characteristics. a**, Output peak signal and idler power as a function of input signal power at λ = 2238 nm, with fixed peak pump power $P_p$ ~ 27.9 W. Measurements show linear amplifier response over a dynamic range larger

than 30 dB, with gain of 23.4 dB and 24.7 dB for signal and idler, respectively. **b**, On-chip signal and idler gain as a function of input peak pump power, with fixed input signal power $P_{sig}$ ~ 0.39 mW at λ = 2238 nm. Transparency is reached for pump power $P_p$ ~ 7.5 W. Onset of gain saturation does not begin until $P_p$ ~ 19 W, at which point the signal and idler gain both exceed 20 dB. **c**, Output peak pump power versus input peak pump power. Experimentally measured data points are shown by the black squares. The blue dash-dotted and red dashed curves illustrate simulated transmission models accounting for nonlinear loss from TPA only and TPA-plus-3PA, respectively.

**Figure 3 | Broadband mid-IR light generation by efficient cascaded FWM.** Transmission spectrum illustrating four orders of cascaded mixing products at the silicon nanophotonic OPA output, covering a total bandwidth of 500 nm. The pulsed pump (labeled peak 1, $P_p$ ~ 21.8 W) is centered at λ = 2180 nm, while the cw signal (labeled peak 2, $P_{sig}$ ~ 4.4 mW) is tuned to λ = 2240 nm. Peaks 3 and 4 correspond to the first-order converted idler and amplified signal, respectively. The hatched region 2086 nm < λ < 2273 nm denotes the spectrum in which on-chip parametric gain is experienced, while net on-chip loss is observed outside this region. Peaks 5-9 are generated outside the on-chip gain spectrum via cascaded mixing processes primarily involving the pulsed pump and the amplified signal and idler. The inset shows the energy diagram for the dominant mixing term contributing to peak 5 at λ = 2055 nm, i.e. non-degenerate FWM of pump pulses from peaks 1 and 3, resulting in wavelength conversion of a signal at peak 4 into an idler at peak 5.

**Figure 1:**

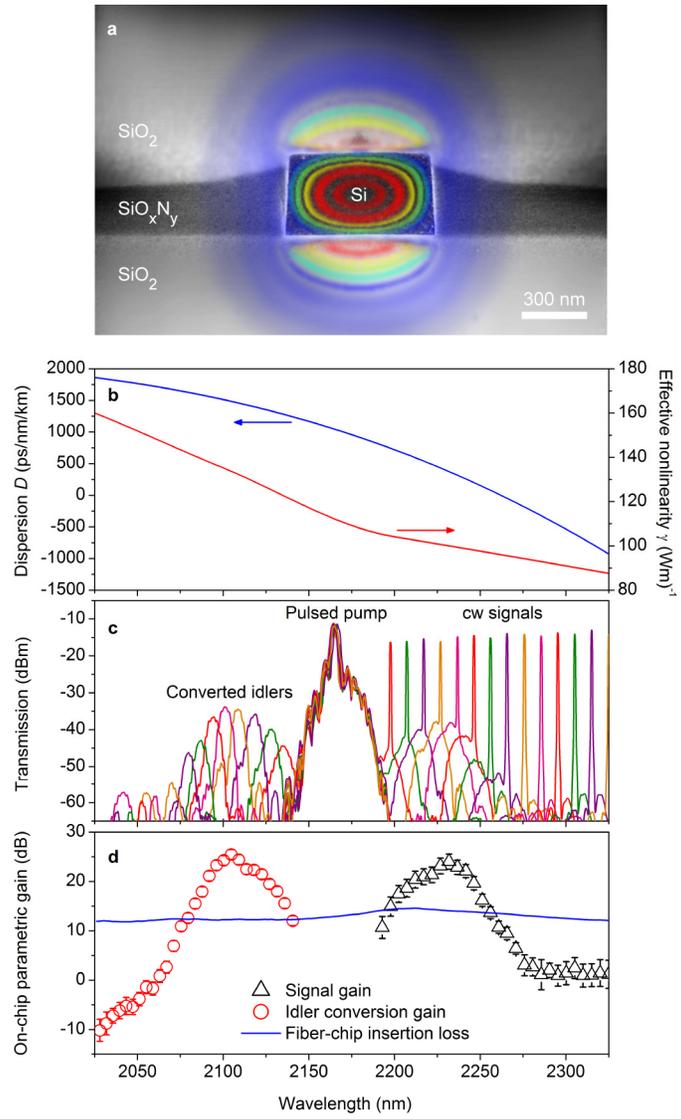

**Figure 2:**

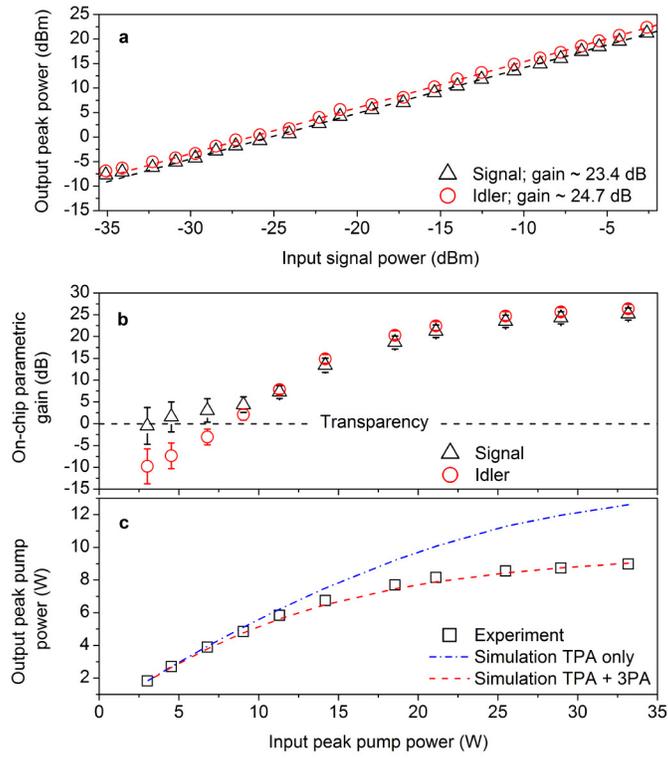

**Figure 3:**

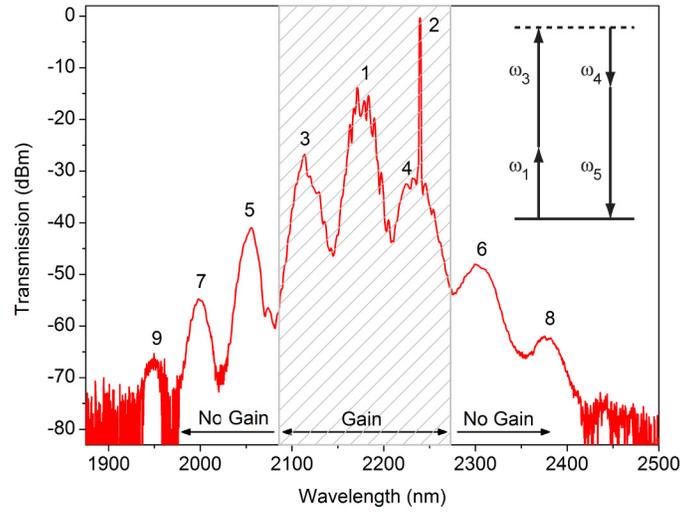

**Supplementary Methods:**

The wavelength-dependent effective index of the silicon nanophotonic waveguide's optical modes were calculated over the range 2.0 μm < λ < 2.5 μm using a commercial finite-element eigen-mode solver (RSoft FemSIM). The material dispersion of Si and $SiO_2$ were incorporated via Sellmeier equations[30], while a constant index of refraction n = 1.56 was assumed for $SiO_xN_y$ (equal to that measured in the visible spectrum), due to a lack of acceptable experimental data. The resulting wavelength-dependent effective index $\bar{n}$ was fitted with a 9$^{th}$-order polynomial and differentiated twice to obtain the dispersion parameter[31] according to $D = -(\lambda/c)(\partial^2 \bar{n}/\partial \lambda^2)$. The same method was applied to the frequency-dependent propagation constant $\beta_0$ to calculate higher order dispersion terms according to $\beta_n = \partial^n \beta_0 / \partial \omega^n$. The effective nonlinearity parameter γ was obtained through the expression $\gamma = 3\omega \operatorname{Re}(\Gamma)/(4\varepsilon_0 A_0 v_g^2)$, where $\omega$ is the optical frequency, $\varepsilon_0$ is the vacuum permittivity, $A_0$ is the area of the Si waveguide core, $v_g$ is the group velocity, and $\Gamma$ is the waveguide effective third-order nonlinear susceptibility. $\Gamma$ was determined by the weighted integral of the third-order susceptibility $\chi^{(3)}$ of bulk silicon over the quasi-TM modal profile, as described elsewhere[5,32]. Experimental values for the TPA coefficient $\beta_{TPA}$ and Kerr coefficient $n_2$ of bulk silicon[13] were used to calculate $\chi^{(3)}$ for wavelengths λ < 2.2 μm. For λ > 2.2 μm, $\beta_{TPA}$ was set to zero, and $n_2$ was fixed to its value at λ = 2.2 μm.

Contour plots of the fundamental quasi-TM mode's zero-dispersion wavelength (ZDWL) and the effective nonlinearity parameter γ at the ZDWL, as a function of the silicon core cross-sectional dimensions (645 nm < width < 735 nm; 380 nm < height < 470 nm) are shown in Supplementary Figures 1 and 2, respectively.

**Supplementary Figures:**

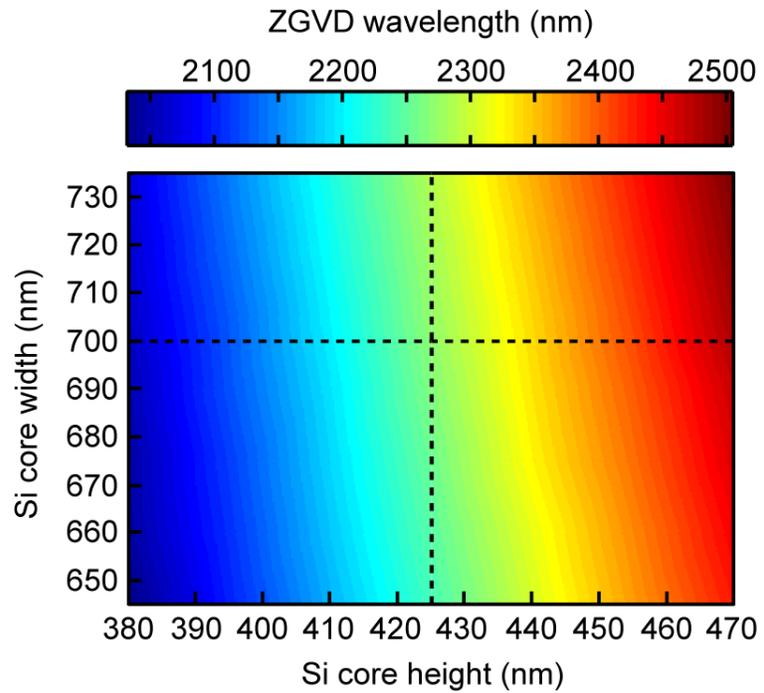

Supplementary Figure 1 | Simulated ZGVD wavelength for the fundamental quasi-TM mode, as a function of the silicon nanophotonic waveguide core cross-sectional dimensions. The intersection of the dotted lines illustrates the design ultimately fabricated, with a silicon core 700 nm wide by 425 nm in height.

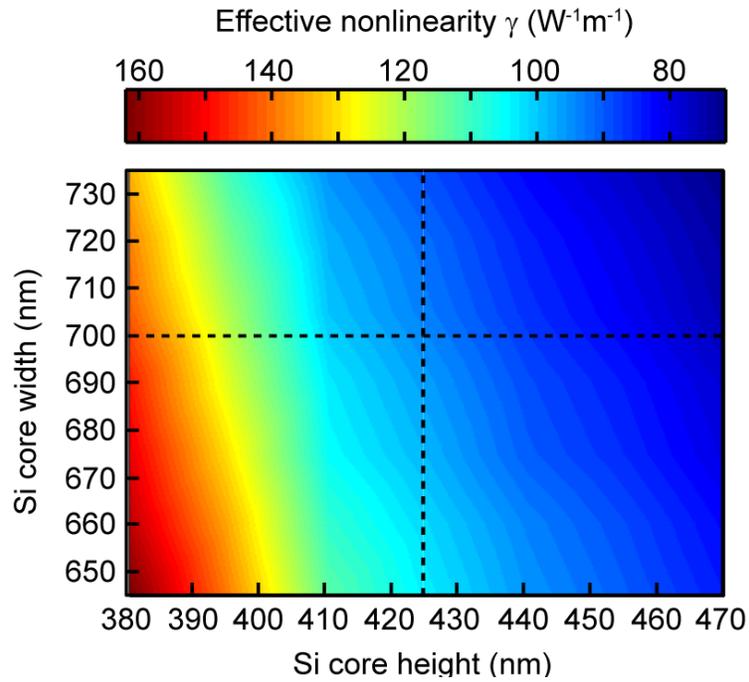

Supplementary Figure 2 | Simulated effective nonlinearity γ (evaluated at the ZDWL) for the fundamental quasi-TM mode, as a function of the silicon nanophotonic waveguide core cross-sectional dimensions. The intersection of the dotted lines illustrates the design ultimately fabricated, with a silicon core 700 nm wide by 425 nm in height.

**Supplementary Notes:**